%% file: main.tex
\documentclass[fleqn,usenatbib]{mnras}

\usepackage{newtxtext,newtxmath}

\usepackage[T1]{fontenc}

\DeclareRobustCommand{\VAN}[3]{#2}
\let\VANthebibliography\thebibliography
\def\thebibliography{\DeclareRobustCommand{\VAN}[3]{##3}\VANthebibliography}

\usepackage{graphicx}	
\usepackage{amsmath}	
\usepackage{orcidlink}
\usepackage{natbib}
\usepackage{booktabs}
\usepackage{multirow}
\usepackage{ulem}

\title[Jax-Accelerated lens modelling]{CSST Strong Lensing Preparation: Fast Modelling of Galaxy-Galaxy Strong Lenses in the Big Data Era}
\author[Cao et al.]{Xiaoyue Cao\orcidlink{0000-0003-4988-9296}$^{1,2}$, 
Ran Li\orcidlink{0000-0003-3899-0612}$^{3,2}$\thanks{E-mail: liran@bnu.edu.cn},
Nan Li\orcidlink{0000-0001-6800-7389}$^{2,1}$\thanks{E-mail: nan.li@nao.cas.cn},
Yun Chen\orcidlink{0000-0001-8919-7409}$^{2,1}$,
Rui Li\orcidlink{0000-0002-3490-4089}$^{4}$,
Huanyuan Shan\orcidlink{0000-0001-8534-837X}$^{5}$,
Tian Li\orcidlink{0009-0005-5008-0381}$^{6}$
\\
$^{1}$School of Astronomy and Space Science, University of Chinese Academy of Sciences, Beijing 100049, China\\
$^{2}$National Astronomical Observatories, Chinese Academy of Sciences, 20A Datun Road, Chaoyang District, Beijing 100012, China\\
$^{3}$ School of Physics and Astronomy, Beijing Normal University,  Beijing 100875, China\\
$^{4}$Institute for Astrophysics, School of Physics, Zhengzhou University, Zhengzhou, 450001, China\\
$^{5}$Shanghai Astronomical Observatory (SHAO), Nandan Road 80, Shanghai 200030, China\\
$^{6}$Institute of Cosmology and Gravitation, University of Portsmouth, Burnaby Rd, Portsmouth PO1 3FX, UK
}

\date{Accepted XXX. Received YYY; in original form ZZZ}

\pubyear{\the\year{}}

\begin{document}
\label{firstpage}
\pagerange{\pageref{firstpage}--\pageref{lastpage}}
\maketitle

\begin{abstract}
Galaxy-galaxy strong lensing provides a powerful probe of galaxy formation, evolution, and the properties of dark matter and dark energy. However, conventional lens-modelling approaches are computationally expensive and require fine-tuning to avoid local optima, rendering them impractical for the hundreds of thousands of lenses expected from surveys such as Euclid, CSST, and Roman Space Telescopes. To overcome these challenges, we introduce \texttt{TinyLensGPU}, a GPU-accelerated lens-modelling tool that employs XLA-based acceleration with JAX and a neural-network-enhanced nested sampling algorithm, \texttt{nautilus-sampler}. Tests on 1,000 simulated galaxy-galaxy lenses demonstrate that on an RTX 4060 Ti GPU, \texttt{TinyLensGPU} achieves likelihood evaluations approximately 2,000 times faster than traditional methods. Moreover, the \texttt{nautilus-sampler} reduces the number of likelihood evaluations by a factor of 3, decreasing the overall modelling time per lens from several days to roughly 3 minutes. Application to 63 SLACS lenses observed by the Hubble Space Telescope recovers Einstein radii consistent with the literature values (within $\lesssim 5\%$ deviation), which is within known systematic uncertainties. Catastrophic failures, where the sampler becomes trapped in local optima, occur in approximately 5\% of the simulated cases and 10\% of the SLACS sample. We argue that such issues are inherent to automated lens modelling but can be mitigated by incorporating prior knowledge from machine learning techniques. This work thus marks a promising step toward the efficient analysis of strong lenses in the era of big data. The code and data are available online\footnotemark.
\end{abstract}
\footnotetext{\url{https://github.com/caoxiaoyue/TinyLensGpu}}



\input{intro.tex}
\input{method.tex}
\input{result.tex}
\input{discuss.tex}
\input{summary.tex}

\section*{Acknowledgements}
We are grateful for the valuable comments and suggestions provided by the referee. This work is supported by the National Key R\&D Program of China (grant Nos. 2022YFF0503403 and 2023YFB3002501), the National Natural Science Foundation of China (Nos. 11988101 and 12473002), the K.C.Wong Education Foundation, and the science research grants from China Manned Space Project with Nos. CMS-CSST-2021-B01 and CMS-CSST-2025-A03. The authors also acknowledge the support of the science research grants from the China Manned Space Project (No. CMS-CSST-2021-A01), the Ministry of Science and Technology of China (No. 2020SKA0110100), and the CAS Project for Young Scientists in Basic Research (No. YSBR-062). XYC acknowledges the support of the National Natural Science Foundation of China (No. 12303006). RL is also supported by the National Natural Science Foundation of China (Nos. 11773032, 12022306). We express our gratitude to ChatGPT, an AI model developed by OpenAI, for its assistance in polishing the English of this paper.

\section*{Data Availability}
The code and data product that supports this work are publicly available from \url{https://github.com/caoxiaoyue/TinyLensGpu}.



\bibliographystyle{mnras}
\bibliography{reference} 




\appendix
\input{appdx_A}

\bsp	
\label{lastpage}
\end{document}

%% file: intro.tex
\section{Introduction}
The light from a background source can be significantly deflected by a massive intervening object on its path to the observer, producing multiple images or distorted arcs of the background source. This phenomenon is known as strong gravitational lensing \citep[SGL,][]{Schneider92, Schneider2006, Meneghetti22}. When both the background source and the lens are galaxies, the phenomenon is termed galaxy-galaxy strong gravitational lensing \citep[GGSL,][]{Bolton06, Treu10, Shu2017, Shajib24_review}. GGSL observations are a powerful tool in modern astronomy, allowing mass estimation of lens galaxies \citep{Bolton08, Shu16_kpc_offset, Shu16, Shajib2021, Etherington2022, Tan2024}, and when combined with photometric and spectroscopic data, enabling studies of their structural evolution \citep{Koopmans09, Auger10, Bolton12, Sonnenfeld2013, Shu2015_s4tm, Chen2019, Li2023PDU, Sheu2024}. Furthermore, GGSL allows the probing of low-mass haloes within lens galaxies or along the line of sight, offering a means to test dark matter models \citep[e.g.,][]{Vegetti2010, Vegetti2012, Li2016, Li2017, He2022_abc, Vegetti2023, Nightingale2024}. In addition, GGSL acts as a cosmic telescope, magnifying high-redshift background sources to reveal their morphology \citep{Newton11, Dye15, Shu16, Ritondale19, Li2024} and dynamical structure \citep{Cheng20, Dye21, Rizzo20, Rizzo21}, which would otherwise remain inaccessible. Moreover, GGSL can constrain cosmological parameters using double Einstein rings \citep{Gavazzi08, Collett14} or time-delay systems \citep{Suyu13, Treu2022, Birrer2024}. Reliable lens modelling, which infers the lens galaxy's mass distribution and reconstructs the intrinsic morphology of the background source, is essential for these applications.

Lens modelling techniques can be broadly classified into two main categories. The first approach employs pre-trained neural networks to directly predict physical parameters from lensing images \citep{Hezaveh2017, Pearson2019, Lemon}. The second is the classical forward modelling approach, which generates synthetic lensing images based on explicit physical models and optimises parameters to match observations \citep[e.g.,][]{Keeton2010, Birrer2015, Nightingale2015}. The first method is computationally efficient, producing results in seconds, but its accuracy depends on the realism of the training data and lacks explicit physical interpretation. Consequently, the classical forward modelling approach is favoured for scientific applications that require detailed and sophisticated lens models.

The classical forward modelling approach for lens modelling presents significant challenges in terms of automation and computational efficiency. These challenges arise from the high dimensionality of the parameter space, which can encompass dozens of parameters, and substantial degeneracies, such as those between the lens ellipticity and the external shear field \citep{Schneider2006}. To mitigate convergence to incorrect local optima, modellers often manually adjust parameter priors based on observed image features, thereby hindering automation \citep[e.g.][]{Bolton08}. Furthermore, exploring these complex parameter spaces is computationally demanding, frequently necessitating millions of forward simulations that can span days or even weeks for a single system \citep{Shajib2021, Etherington2022, Schmidt2022, Tan2024}.

GPU acceleration significantly improves classical forward modelling by taking advantage of its efficiency in array computations, which are fundamental to lensing calculations. Early software, such as \texttt{Lensed} \citep{Tessore2016}, used OpenCL to accelerate lens modelling computations on GPUs. Recently, frameworks such as JAX\footnote{\url{https://github.com/jax-ml/jax}} \citep{jax2018github}, an open-source library developed by Google, have emerged as powerful tools for GPU-accelerated computations. JAX compiles Python code through the Accelerated Linear Algebra (XLA) library, enabling highly efficient GPU execution. Additionally, JAX's automatic differentiation capability \citep{autodiff} improves sampling efficiency and enhances its ability to fit complex models \citep{Chianese2020}.

Pioneering lens modelling software packages, such as \texttt{Gigalens} \citep{Gu2022} and \texttt{Herculens} \citep{Galan2022}, have been developed using JAX. However, substantial opportunities remain for further improvement. For example, \texttt{Gigalens} employs a gradient-based optimiser \citep{autodiff_optimizer} combined with variational inference \citep{VI} and Hamiltonian Monte Carlo (HMC) sampling \citep{HMC} to explore model solutions, a process that typically requires millions of forward simulations. More efficient sampling methods could potentially reduce the number of forward simulations needed to explore the parameter space. \texttt{Herculens} utilises automatic differentiation to fit complex models with many free parameters. However, the current unbatched code structure of \texttt{Herculens} limits the full utilisation of GPU computational resources.

Ongoing and upcoming sky surveys, including the Euclid Space Telescope \citep{euclid_survey}, the China Space Station Telescope (CSST) \citep{CSST_zhanhu}, and the Nancy Grace Roman Space Telescope \citep{Roman}, are expected to identify hundreds of thousands of GGSL systems \citep{Collett15, Cao2024, Weiner2020}. This represents an increase of nearly three orders of magnitude compared to current samples. Existing lens modelling frameworks, such as \texttt{Lenstronomy} \citep{Lenstronomy} and \texttt{PyAutoLens} \citep{Pyautolens}, lack the computational efficiency required to process this unprecedented volume of data.

In this work, we develop a new lens modelling code---\texttt{TinyLensGpu}, which leverages JAX and GPU acceleration for forward simulation computations. Our program batches the computation of model lensing images into multidimensional arrays, effectively utilising GPUs for large-scale array processing. Furthermore, we employ a nested sampler named \texttt{nautilus-sampler}\footnote{\url{https://github.com/johannesulf/nautilus}} \citep{nautilus}, which utilises neural networks to propose new sample points, enhancing the efficiency of lens modelling sampling.

This paper is organised as follows: Section~\ref{sec:method} outlines the methodology underlying \texttt{TinyLensGpu}. Section~\ref{sec:result} presents the main results. In Section~\ref{sec:res_mock}, We evaluate the performance of \texttt{TinyLensGpu} using mock lensing data with image quality comparable to that expected from ongoing and upcoming space-telescope surveys such as Euclid, CSST, and Roman. Subsequently, in Section~\ref{sec:res_real}, we apply our code to 63 real lenses acquired by the \textit{Hubble Space Telescope} (HST) for additional validation. Finally, Section~\ref{sec:discuss} discusses the insights into automatic lens modelling gained from analysing 1,000 mock lenses and 63 real HST lenses, as well as the future prospects of this work. Our conclusions are summarised in Section~\ref{sec:summary}. Throughout this study, we adopt a flat $\Lambda$CDM cosmology with the parameters $\Omega_m = 0.3$, $\Omega_\Lambda = 0.7$, and $H_0 = 70\, \mathrm{km}\,\mathrm{s}^{-1}\,\mathrm{Mpc}^{-1}$. All codes and relevant data used in this study are publicly available at the following link: \url{https://github.com/caoxiaoyue/TinyLensGpu}.

%% file: method.tex
\section{Methodology}
\label{sec:method}
A typical galaxy-galaxy strong lensing image consists of a bright central blob produced by the lens galaxy, surrounded by an extended lensed arc or multiple images representing the distorted light from the source. The goal of galaxy-galaxy strong lens modelling is to reconstruct the mass distribution of the lens galaxy and the intrinsic brightness distribution of the source galaxy before lensing, by fitting pixel-wise brightness values in the observed image. Section~\ref{sec:lensing_theory} outlines the basic lensing physics, and demonstrates forward simulations to generate model lensing images. Section~\ref{sec:prob_framework} introduces the Bayesian statistical framework for parameter inference in lens modelling. Section~\ref{sec:linear_solving} describes the inversion framework employed to solve for linear parameters using matrix notation. Section~\ref{sec:light_mass_models} presents the physical models that characterise the mass and light distributions of the galaxy, as implemented in the \texttt{TinyLensGpu} code.

\subsection{Lensing Theory and Forward Simulation}
\label{sec:lensing_theory}
In galaxy-galaxy strong gravitational lensing, the physical sizes of both the lensing object and the luminous source are negligible compared to their distances from the observer and each other. Consequently, their mass and brightness distributions are projected onto two dimensions via the thin-lens approximation \citep{Meneghetti22}, yielding the lens and source planes corresponding to the lens and source galaxies, respectively. The positions $\boldsymbol{\theta}$ on the lens plane and $\boldsymbol{\beta}$ on the source plane are related by the lens equation:
\begin{equation}
\label{eq:lens_mapping}
\boldsymbol{\beta}(\boldsymbol{\theta})=\boldsymbol{\theta}-\boldsymbol{\alpha}(\boldsymbol{\theta}).
\end{equation}
$\boldsymbol{\alpha}(\boldsymbol{\theta})$ denotes the deflection angle at position $\boldsymbol{\theta}$, determined by the mass distribution of the lens galaxy. Let $\mathbf{I_\mathrm{L}}$ and $\mathbf{I_\mathrm{S}}$ represent the surface brightness distributions of the lens galaxy and the source galaxy, respectively. The surface brightness $\mathbf{I}(\boldsymbol{\theta})$ at any position $\boldsymbol{\theta}$ on the ideal lensing image is expressed as
\begin{equation}
\label{eq:image_forming}
\begin{aligned}
\mathbf{I}(\boldsymbol{\theta}) & =\mathbf{I}_{\mathrm{L}}(\boldsymbol{\theta})+\mathbf{I}_{\mathrm{S}}(\boldsymbol{\beta}(\boldsymbol{\theta})) \\
& =\mathbf{I}_{\mathrm{L}}(\boldsymbol{\theta})+\mathbf{I}_{\mathrm{S}}(\boldsymbol{\theta}-\boldsymbol{\alpha}(\boldsymbol{\theta})).
\end{aligned}
\end{equation}
This equation holds because gravitational lensing conserves surface brightness. Approximating a pixel's average surface brightness using its central value is inaccurate, especially for sources with steep spatial brightness gradients. To calculate model lensing images, $4\times4$ supersampling is first performed, followed by binning the model image to restore the original resolution. At this stage, the model lensing image does not account for blurring effects caused by the telescope's diffraction limit or atmospheric turbulence. These blurring effects are quantitatively described by the point spread function (PSF). The PSF characterises how an ideal point source spreads into a spot and is typically represented by a two-dimensional kernel in lens modelling. Let $\mathbf{K}$ represent the PSF kernel; the blurred lensing image $\mathbf{I}^\mathrm{m}$ is then related to the ideal lensing image $\mathbf{I}(\boldsymbol{\theta})$ through the convolution operation:
\begin{equation}
\label{eq:image_convolution}
\mathbf{I}^{\mathrm{m}}(\boldsymbol{\theta})=\int \mathbf{K}\left(\boldsymbol{\theta}-\boldsymbol{\theta}^{\prime}\right) \mathbf{I}\left(\boldsymbol{\theta}^{\prime}\right) d \boldsymbol{\theta}^{\prime}.
\end{equation}
$\mathbf{I}^\mathrm{m}$ is the final output of the forward simulation and is compared with the observed lensing image to constrain the lens model parameters.

\subsection{Bayesian Probability Framework}
\label{sec:prob_framework}
We denote the model parameters associated with the lens galaxy as $\vec{\eta}$. This vector comprises the parameters describing both the brightness distribution, $\vec{\eta}_l$, and the mass distribution, $\vec{\eta}_m$, such that $\vec{\eta} = \{\vec{\eta}_l, \vec{\eta}_m\}$. The parameters describing the source brightness distribution are represented by the vector $\vec{s}$. Consequently, the complete set of lens model parameters is given by $\vec{\xi} = \{\vec{\eta}, \vec{s}\}$. According to Bayes' theorem, the posterior probability distribution of $\vec{\xi}$ given the lensing image $\vec{d}$, denoted as $P(\vec{\xi} \mid \vec{d})$, can be expressed as follows:
\begin{equation}
P(\vec{\xi} \mid \vec{d}) = \frac{P(\vec{d} \mid \vec{\xi}) P(\vec{\xi})}{P(\vec{d})}.
\end{equation}
Here, the prior probability distribution $P(\vec{\xi})$ encodes the modeller's prior knowledge of the parameter values. The term $P(\vec{d} \mid \vec{\xi})$ is the likelihood function, quantifying the probability of observing the lensing image given the model parameters $\vec{\xi}$. The denominator, $P(\vec{d}) = \int P(\vec{d} \mid \vec{\xi}) P(\vec{\xi}) \, d\vec{\xi}$, represents the marginal likelihood, also known as Bayesian evidence, which can be used for model comparison. For typical optical-band lensing images, the likelihood function is often approximated by a Gaussian distribution:
\begin{equation}
P(\vec{d} \mid \vec{\xi}) = \prod_i \frac{1}{\sqrt{2\pi\sigma_i^2}} \exp\left(-\frac{1}{2} \frac{\left(d_i - I^\mathrm{m}_i(\vec{\xi})\right)^2}{\sigma_i^2}\right),
\end{equation}
where $d_i$ denotes the observed brightness at the $i$-th pixel, $\sigma_i$ represents the corresponding noise level, and $I^\mathrm{m}_i(\vec{\xi})$ is the model-predicted surface brightness at pixel $i$.

The lens modelling parameters, $\vec{\xi}$, are categorised into linear ($\vec{\xi}_l$) and non-linear ($\vec{\xi}_n$) components based on the method used to determine their optimal values. The non-linear parameters, $\vec{\xi}_n$, are determined through non-linear search algorithms such as nested sampling \citep{nested_sampling}. For a given set of non-linear parameters $\vec{\xi}_n$ and $\vec{d}$, the optimal linear parameters, denoted by $\vec{\xi}_l^B$, can be directly computed using a linear least-squares algorithm (see Section~\ref{sec:linear_solving} for details). Since the likelihood function, $P(\vec{d} \mid \vec{\xi})$, is sharply peaked around the optimal linear parameters, it can be approximated by a delta function. Consequently, the likelihood function is simplified to
\begin{equation}
\begin{aligned}
P(\vec{d}\mid \vec{\xi}) &= \int P(\vec{d} \mid \vec{\xi}_n, \vec{\xi}_l)P(\vec{\xi}_l)d\vec{\xi}_l \\
&= P(\vec{d}\mid \vec{\xi}_n,\vec{\xi}_l^B).
\end{aligned}
\end{equation}
Therefore, by decomposing the lens modeling parameters into linear and non-linear components and analytically computing the optimal linear parameters, we reduce the dimensionality of the non-linear parameter space, thereby improving the efficiency of the non-linear search.

\subsection{Linear Inversion Using Matrix Notation}
\label{sec:linear_solving}
In Section~\ref{sec:prob_framework}, we mentioned that the linear lens modelling parameters can be analytically computed using the linear least-squares algorithm, also known as linear inversion, given a set of non-linear parameters. This reduces the dimensionality of the non-linear parameter space and enhances fitting efficiency. We now elaborate on the formal mathematical framework of this linear inversion using matrix notation.

Suppose the lensing image being modelled consists of $N_d$ pixels and is represented by the column vector $\vec{d}$, where each entry corresponds to the brightness value at a given pixel. Similarly, the noise is represented by the column vector $\vec{n}$. The brightness distributions of the lens and source galaxies are modelled as superpositions of $N_l$ and $N_s$ light profiles, respectively. The forward simulation process is then expressed as
\begin{equation}
\begin{aligned} 
\vec{d} & = \boldsymbol{B}\,\boldsymbol{L_l}\,\vec{s_l} + \boldsymbol{B}\,\boldsymbol{L_s}\,\vec{s_s} + \vec{n} \\ 
& = \boldsymbol{B}\,\bigl[\boldsymbol{L_l} \mid \boldsymbol{L_s}\bigr] \begin{pmatrix} \vec{s_l} \\ \vec{s_s} \end{pmatrix} + \vec{n} \\ 
& = \boldsymbol{B}\,\boldsymbol{L}\,\vec{s} + \vec{n} \\ 
& = \boldsymbol{M}\,\vec{s} + \vec{n}, 
\end{aligned} 
\end{equation} 
where:
\begin{itemize}
    \item $\vec{s_l}$ is a column vector of length $N_l$ that contains the amplitudes of the lens light profiles. Similarly, $\vec{s_s}$ is a column vector of length $N_s$ that represents the amplitudes of the source light profiles. 
    \item $\boldsymbol{L_l}$ is an $N_d \times N_l$ matrix, where each column corresponds to the lensing image of an individual lens light profile with unit amplitude. The entries of $\boldsymbol{L_l}$ depend on the non-linear parameters of the lens light profiles, including their centres and ellipticities.
    \item $\boldsymbol{L_s}$ is an $N_d \times N_s$ matrix, where each column corresponds to the lensed image of an individual source light profile with unit amplitude. The entries of $\boldsymbol{L_s}$ depend on the non-linear parameters of the source light profiles as well as the mass model parameters of the lens galaxy.
    \item $\boldsymbol{B}$ is an $N_d \times N_d$ matrix that represents the blurring effect of the PSF.
    \item $\boldsymbol{L}$ is an $N_d \times (N_l + N_s)$ matrix formed by concatenating $\boldsymbol{L_l}$ and $\boldsymbol{L_s}$ column-wise: $\boldsymbol{L} = \bigl[\boldsymbol{L_l} \mid \boldsymbol{L_s}\bigr]$. The combined amplitude vector $\vec{s}$ is then:
    \[
    \vec{s} = \begin{pmatrix} \vec{s_l} \\ \vec{s_s} \end{pmatrix}.
    \]
    \item The matrix $\boldsymbol{M}$ is defined as $\boldsymbol{M} \equiv \boldsymbol{B}\,\boldsymbol{L}$.
\end{itemize}
The lensing image $\vec{d}$ and the linear parameters $\vec{s}$ are related through a linear response, given a set of nonlinear parameters encoded in the matrix $\boldsymbol{M}$. Consequently, determining the optimal $\vec{s}$ requires minimizing the following penalty function: 
\[ \|(\boldsymbol{M}\,\vec{s} - \vec{d})/\vec{n}\|, \]  
subject to the constraint $\vec{s} \geq 0$. This constraint arises from the physical requirement that light profile amplitudes must be non-negative. The minimisation defines a standard non-negative linear least-squares problem\footnote{In the bottom panel of Figure~\ref{fig:slacs_example}, we show that an unphysical model reconstruction may occur if the amplitude of the light profiles is not estimated using non-negative linear least squares.} \citep{NNLS}. We solve it using the iterative algorithm described in \cite{FNNLS}\footnote{Our implementation is based on a JAX port of the Python package \texttt{fnnls}: \url{https://github.com/jvendrow/fnnls}.}, which yields an analytical solution for the optimal $\vec{s}$.

\subsection{Light And Mass Distribution Models}
\label{sec:light_mass_models}
The Sérsic profile \citep{Sersic} is commonly used to describe a galaxy's light distribution and is defined as
\begin{equation}
I(R) = I_{\mathrm{e}} \exp \left\{-b_{\mathrm{n}}\left[\left(\frac{R}{r_{\mathrm{e}}}\right)^{1/n} - 1\right]\right\},
\end{equation}
where $r_{\mathrm{e}}$ is the effective radius of the galaxy, and $I_{\mathrm{e}}$ is the surface brightness at that radius. The Sérsic index $n$ characterises the concentration of the light distribution. The coefficient $b_{\mathrm{n}}$ depends only on $n$ and satisfies
\begin{equation}
\Gamma(2n) = 2 \gamma\left(2n, b_{\mathrm{n}}\right),
\end{equation}
which ensures that the flux enclosed within $r_{\mathrm{e}}$ is exactly half of the total galaxy flux. Here, $\Gamma$ and $\gamma$ denote the complete and incomplete gamma functions, respectively. The one-dimensional Sérsic profile extends to a two-dimensional brightness distribution by applying the following elliptical coordinate transformation, replacing $R$ with $R_p$:
\begin{equation}
\begin{aligned}
x_{1,p} & = \left(x_1 - c_{1,p}\right) \cos \theta_p + \left(x_2 - c_{2,p}\right) \sin \theta_p, \\
x_{2,p} & = -\left(x_1 - c_{1,p}\right) \sin \theta_p + \left(x_2 - c_{2,p}\right) \cos \theta_p, \\
R_p & = \sqrt{x_{1,p}^2 q_p + \frac{x_{2,p}^2}{q_p}},
\end{aligned}
\end{equation}
where $\boldsymbol{x} = (x_1, x_2)$ represents the position in Cartesian coordinates, $(c_{1,p}, c_{2,p})$ denotes the center of the light distribution, and $\theta_p$ and $q_p$ denote the position angle and axial ratio (defined as the ratio of the minor to major axes) of the elliptical model, respectively. For enhanced sampling efficiency in lens modeling, $\theta_p$ and $q_p$ are often reparameterized in terms of ellipticities \citep{Birrer2015}:
\begin{equation}
\left(\epsilon_1, \epsilon_2\right) = \frac{1-q_p}{1+q_p}(\cos(2\theta_p), \sin(2\theta_p)),
\end{equation}
where $\epsilon_1$ and $\epsilon_2$ are the Cartesian components of the ellipticity.

The mass profile of the lens galaxy is well described by a power-law model:  
\begin{equation}
\kappa(R) = \frac{3 - \gamma}{2} \left( \frac{\theta_{\mathrm{E}}}{R} \right)^{\gamma - 1},
\end{equation}
where $\theta_{\mathrm{E}}$ denotes the Einstein radius, and $\gamma$ is the power-law slope. Similar to the light profile, this mass profile can be extended to a two-dimensional elliptical distribution, known as the elliptical power-law (EPL) model \citep{EPL_model}. When $\gamma = 2$, the EPL model simplifies to the singular isothermal ellipsoid (SIE) model \citep{Kormann1994}. The lens galaxy’s neighbouring galaxies or cosmic structures along the line of sight can induce external shear \citep{Keeton1997}. The external shear's lensing potential, expressed in polar coordinates $(\theta, \varphi)$, is defined as
\begin{equation}
\psi^{\mathrm{ext}}(\theta, \varphi) = -\frac{1}{2} \gamma^{\mathrm{ext}} \theta^2 \cos 2\left(\varphi - \phi^{\mathrm{ext}}\right),
\end{equation}
where $\gamma^{\mathrm{ext}}$ and $\phi^{\mathrm{ext}}$ represent the shear magnitude and orientation angle, respectively. In lens modelling, the external shear is often reparameterized in terms of Cartesian components $(\gamma_1^{\mathrm{ext}}, \gamma_2^{\mathrm{ext}})$, where
\begin{equation}
\gamma^{\mathrm{ext}} = \sqrt{\left(\gamma_1^{\mathrm{ext}}\right)^2 + \left(\gamma_2^{\mathrm{ext}}\right)^2}, \quad \tan 2\phi^{\mathrm{ext}} = \frac{\gamma_2^{\mathrm{ext}}}{\gamma_1^{\mathrm{ext}}}.
\end{equation}

%% file: result.tex
\section{Result}
\label{sec:result}
This section systematically evaluates the performance of \texttt{TinyLensGpu}, with a particular focus on employing a uniform modelling pipeline to process a large number of lenses. Initially, we apply \texttt{TinyLensGpu} to a sample of 1000 mock lenses, as detailed in Section~\ref{sec:res_mock}. These mock lenses are designed with image properties that mimic those anticipated in upcoming space telescope surveys, such as CSST, Euclid, and Roman. This initial test aims to assess the capacity of \texttt{TinyLensGpu} to rapidly and robustly model large lens samples, specifically by demonstrating its ability to stably identify true solutions without converging to local minima in the parameter space. Subsequently, to address the complexities inherent in real lensing observations, which are not fully represented in our mock datasets—such as the irregular morphologies of the lens and source galaxies—we also evaluate the performance of \texttt{TinyLensGpu} using 63 real lenses imaged by the HST \citep{Bolton08}, as described in Section~\ref{sec:res_real}. By comparing our results with those reported in prior studies, we confirm that \texttt{TinyLensGpu} can automatically model large lens samples both efficiently and reliably. The default prior settings for the various mass and light models used in this work are summarised in Table~\ref{tab:prior_setting}.

\begin{table}
    \centering
    \renewcommand{\arraystretch}{1.2}
    \begin{tabular}{llc}
        \toprule
        \textbf{Model Component} & \textbf{Parameter} & \textbf{Prior} \\ 
        \midrule
        \multirow{4}{*}{SIE} 
        & $\theta_E$ & $\mathcal{U}(0.0, 3.5)$ \\
        & $e_1$ & $\mathcal{N}(0.0, 0.3)$ \\
        & $e_2$ & $\mathcal{N}(0.0, 0.3)$ \\
        & $(x, y)$ & $\mathcal{N}(0.0, 0.1)$ \\
        \midrule
        \multirow{2}{*}{External Shear} 
        & $\gamma_1$ & $\mathcal{U}(-0.2, 0.2)$ \\
        & $\gamma_2$ & $\mathcal{U}(-0.2, 0.2)$ \\
        \midrule
        \multirow{6}{*}{Sersic (Source)} 
        & $r_e$ & $\mathcal{U}(0.0, 2.0)$ \\
        & $n$ & $\mathcal{U}(0.3, 2.5)$ \\
        & $e_1$ & $\mathcal{N}(0.0, 0.3)$ \\
        & $e_2$ & $\mathcal{N}(0.0, 0.3)$ \\
        & $(x, y)$ & $\mathcal{N}(0.0, 1.0)$ \\
        & $I_e$ & Fixed (solved linearly) \\
        \midrule
        \multirow{5}{*}{Sersic (Lens)} 
        & $r_e$ & $\mathcal{U}(0.0, 4.0)$ \\
        & $n$ & $\mathcal{N}(4.0, 1.0)$ \\
        & $e_1$ & $\mathcal{N}(0.0, 0.3)$ \\
        & $e_2$ & $\mathcal{N}(0.0, 0.3)$ \\
        & $(x, y)$ & Fixed (to lens mass center) \\
        & $I_e$ & Fixed (solved linearly) \\
        \bottomrule
    \end{tabular}
    \caption{Default parameter settings used in this work unless stated otherwise. Here, $\mathcal{U}(a, b)$ represents a uniform prior with lower and upper bounds $a$ and $b$, respectively, while $\mathcal{N}(c, d)$ represents a Gaussian prior with mean $c$ and standard deviation $d$.}
    \label{tab:prior_setting}
\end{table}

\subsection{Testing on Mock Datasets}
\label{sec:res_mock}
\begin{figure*}
	\includegraphics[width=\textwidth]{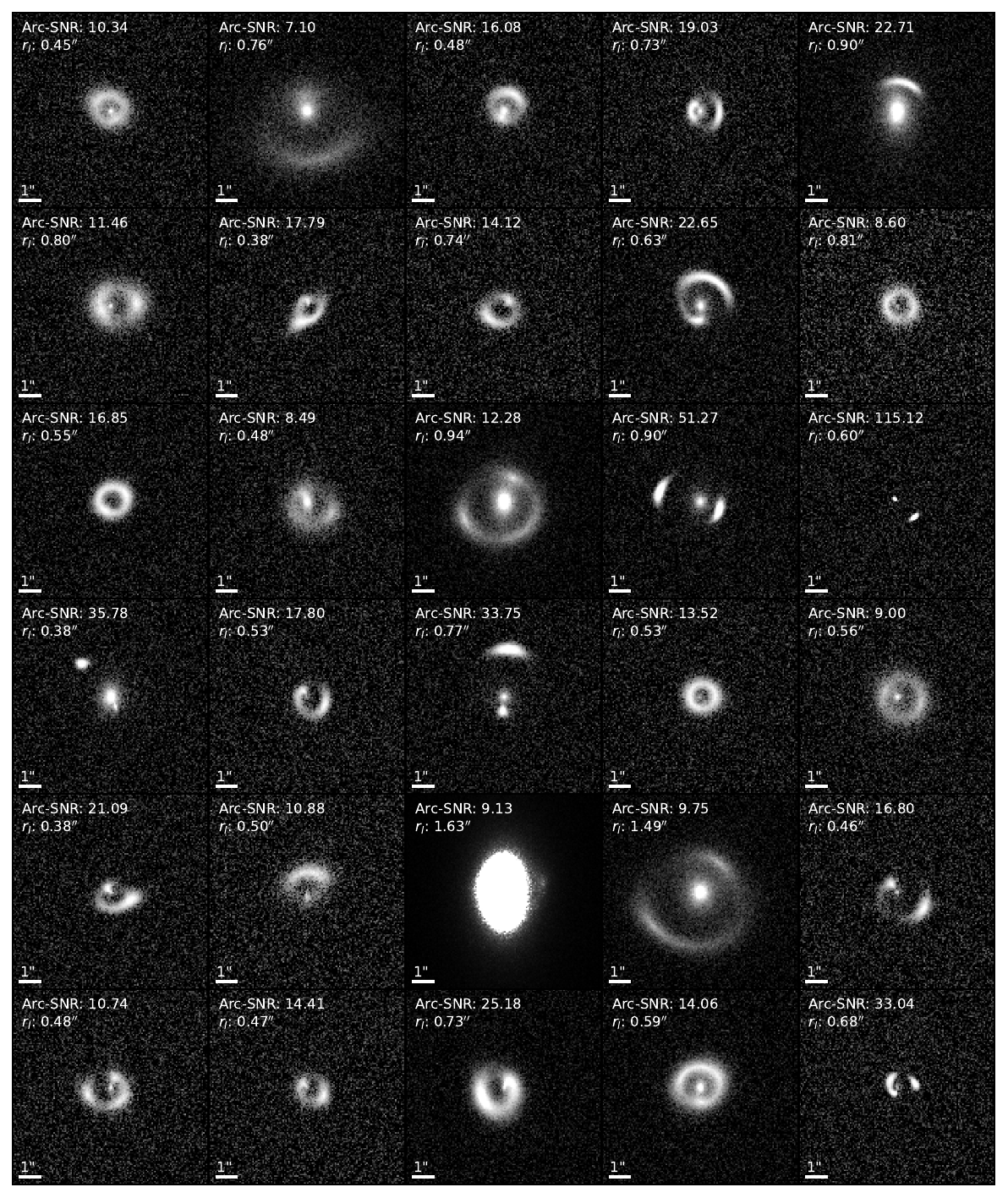}
        \caption{Image atlas of 30 randomly selected systems from a set of 1,000 simulated lenses with CSST-like image quality. The maximum signal-to-noise ratio of the lensed arc and the half-light radius of the lens galaxy are indicated in the top-left corner. The lens galaxy may appear small or absent due to its faintness in the $g$-band.}
    \label{fig:csst_atlas}
\end{figure*}

\begin{figure*}
	\includegraphics[width=\textwidth]{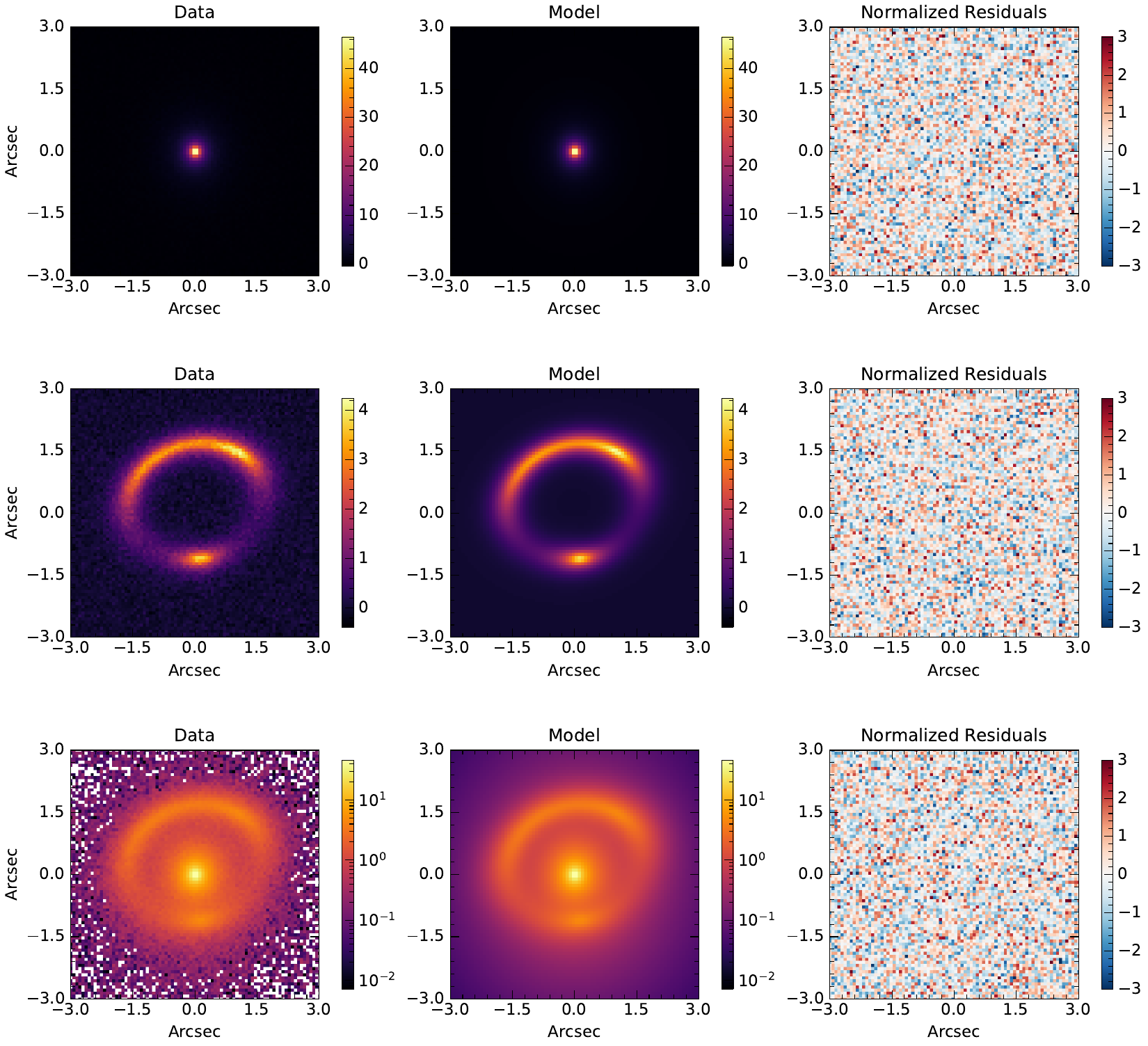}
        \caption{Lens modelling results for mock systems illustrating the basic capabilities of \texttt{TinyLensGpu}. From top to bottom, the rows correspond to lens modelling results for datasets containing: (i) only the lens light, (ii) only the lensed source light, and (iii) both the lens and source light. From left to right, the columns represent the data image, the model image, and the normalised residuals, defined as the difference between the data and model, divided by the image noise. The data and model images in the bottom row are presented on a logarithmic scale to enhance the visibility of the faint lensed arc, where white pixels represent small negative values caused by sky background noise.}
    \label{fig:demo_cases}
\end{figure*}

\begin{figure*}
	\includegraphics[width=\textwidth]{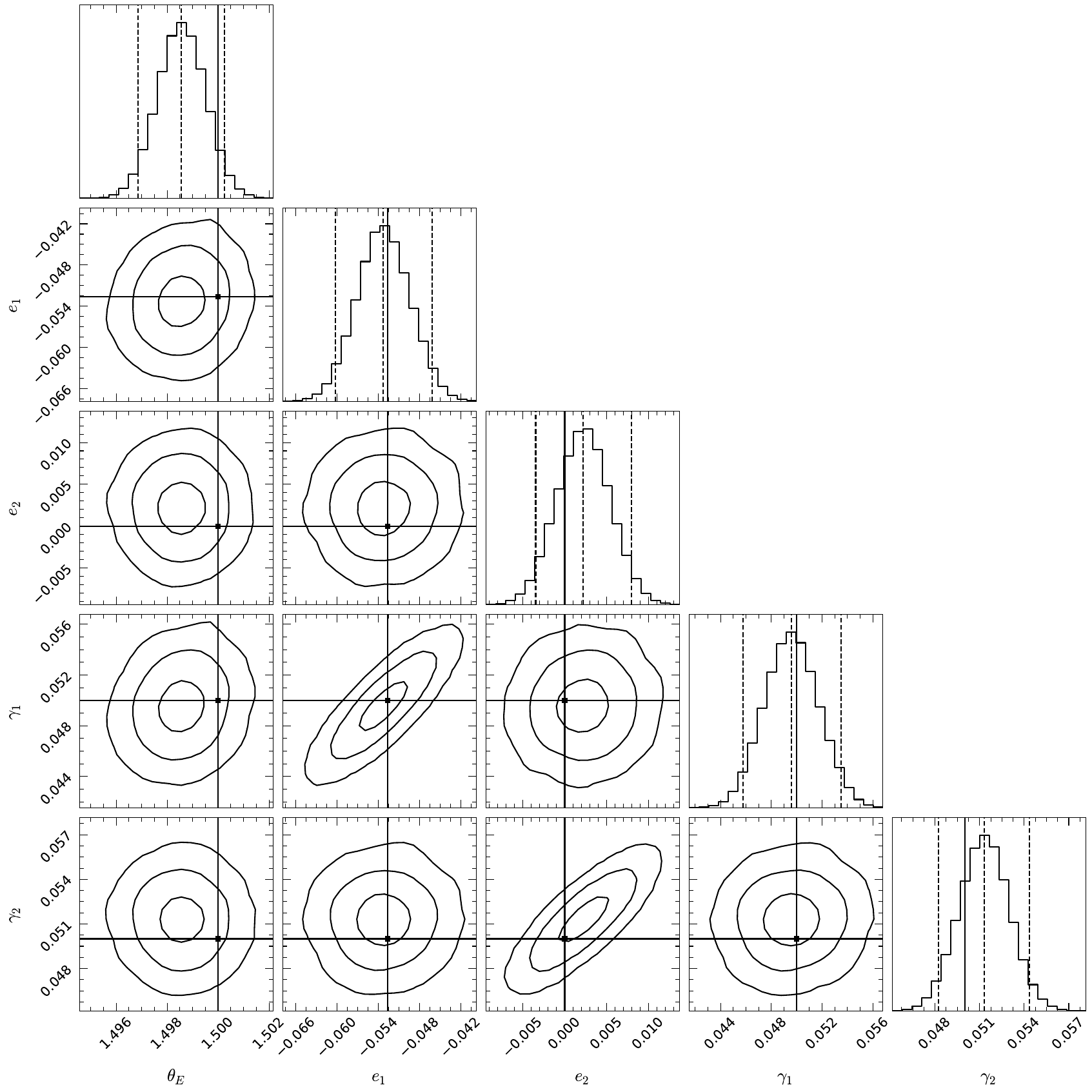}
        \caption{Posterior distributions of key lens mass parameters for the example lens system shown in the bottom row of Figure~\ref{fig:demo_cases}. The parameters include the Einstein radius ($\theta_E$), the ellipticity ($e_1$ and $e_2$) of the main lens, and the external shear ($\gamma_1$ and $\gamma_2$). Contours represent the 68.27\%, 95.45\%, and 99.73\% credible intervals. Dashed lines indicate the median and 95.45\% credible interval, while solid lines denote the ground truth values.}
    \label{fig:demo_posterior}
\end{figure*}

\begin{figure*}
	\includegraphics[width=\textwidth]{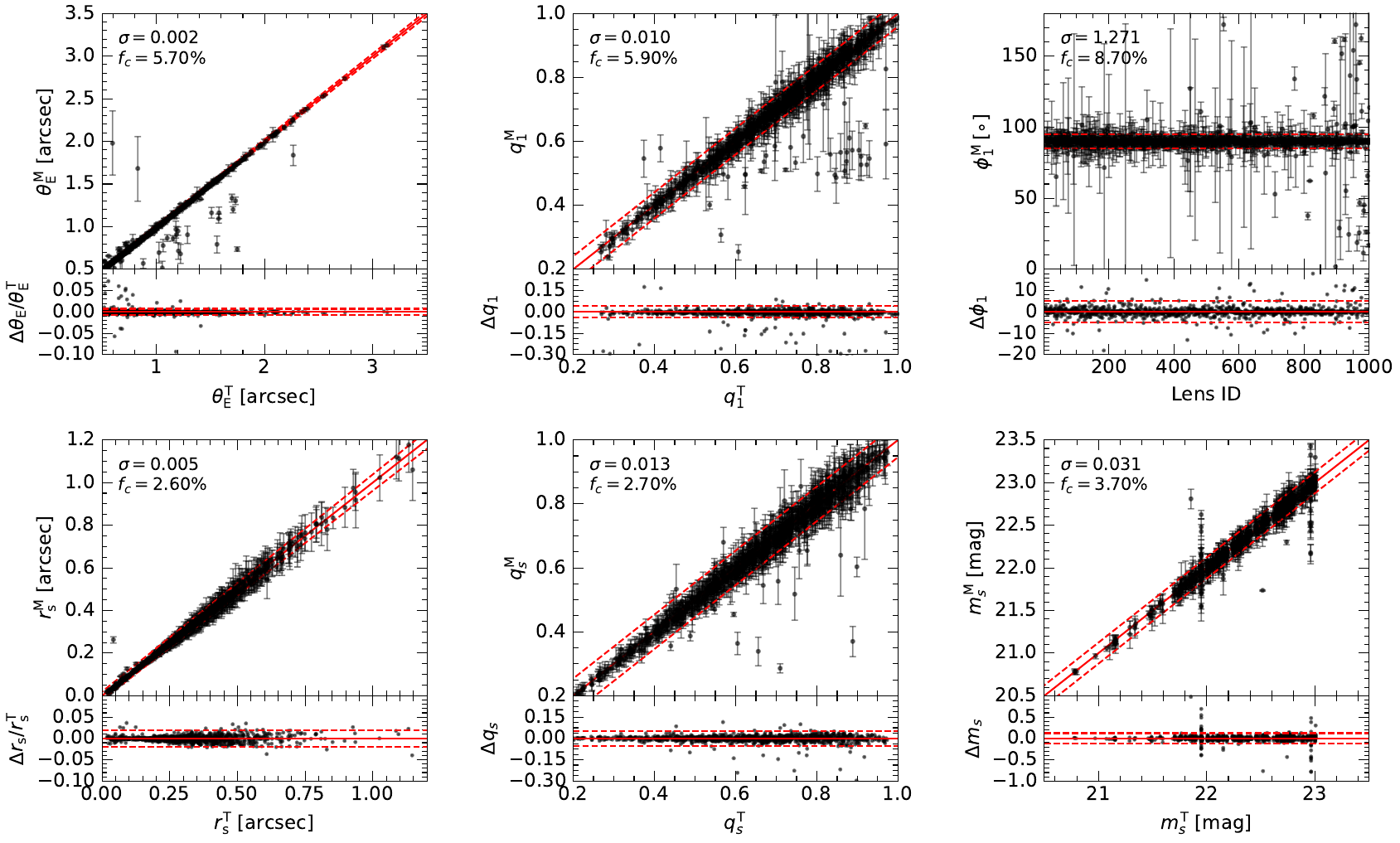}
        \caption{Comparison of lens modelling results from \texttt{TinyLensGpu} for 1000 simulated galaxy-galaxy strong lenses with CSST-like image quality against their ground-truth values. The panels present parameters of the lens galaxy (Einstein radius $\theta_E$, axis ratio $q_1$, and position angle $\phi_l$) and the source galaxy (half-light radius $r_s$, axis ratio $q_s$, and unlensed magnitude $m_s$). In each panel, the top subpanel compares the modelled values (denoted by the superscript $\rm{M}$) with the ground-truth values (denoted by the superscript $\rm{T}$). The bottom subpanel illustrates the scatter between these values, expressed as relative deviations for $\theta_E$ and $r_s$ and absolute deviations for the remaining parameters. The typical scatter ($\sigma$), computed using the Normalised Median Absolute Deviation, is indicated in the top subpanels. The red solid line represents the 1:1 relation, while the dashed lines delineate the region within $4\sigma$. The fraction of modelled samples that fall outside this $4\sigma$ region ("catastrophic outliers") is denoted by $f_c$. The error bars represent the $3\sigma$ ($99.73\%$) credible interval. For the position angle of the lens galaxy, $\phi_l$, where the ground-truth values of all mock lenses are set to zero, the x-axis of the top-right panel represents the mock lens ID. Systems with an axis ratio close to 1 exhibit large measurement uncertainties in $\phi_l$, as it is effectively unconstrained in these cases.}
    \label{fig:mock_summary}
\end{figure*}

\subsubsection{Mock Lens Creation}
\label{sec:gen_mock}
We utilise the mock lens catalogue from \cite{Cao2024} to simulate 1,000 galaxy-galaxy strong lenses anticipated to be detected in future CSST surveys. The catalogue is generated by first populating the sky with early-type galaxies—the dominant deflectors in galaxy-galaxy strong lensing—using empirical population relations derived from observational data. Source galaxies are then incorporated, with their properties drawn from a mock catalogue based on cosmological simulations. The simulation identifies instances where source galaxies are strongly lensed by foreground early-type galaxies and assesses whether the resulting strong lenses are detectable by CSST. In our simulations, the lens galaxy's light distribution is modelled using an elliptical de Vaucouleurs profile (i.e., a S\'ersic profile with $n=4$), and its mass distribution is modelled using a Singular Isothermal Ellipsoid. The source galaxy’s light distribution follows an elliptical S\'ersic profile. Figure~\ref{fig:csst_atlas} presents lensing images of 30 randomly selected systems from the 1,000 simulated mock lenses.

\subsubsection{Example Use Cases}
\label{sec:demo}
Figure~\ref{fig:demo_cases} presents the modelling results for three simulated systems, demonstrating the versatility of \texttt{TinyLensGpu}. In the top-left panel, where no light from the background source is visible, \texttt{TinyLensGpu} models the foreground galaxy’s light distribution using S\'ersic profiles (top-middle panel), producing residuals at the noise level (top-right panel). This functionality is analogous to that of conventional galaxy photometry software such as \texttt{galfit} \citep{Peng2002}. When the lens galaxy’s light is too faint to detect and only the lensed arc of the background source is observed, \texttt{TinyLensGpu} focuses on the lensed arc to reconstruct both the lens galaxy’s mass distribution and the intrinsic light distribution of the source, as shown in the middle row. In the most general case, where both the foreground lens and the background source are visible (bottom row), \texttt{TinyLensGpu} simultaneously models both components, effectively separating their light contributions and reconstructing the lens galaxy’s mass distribution along with the source’s intrinsic light distribution. All model parameters are recovered without bias, as demonstrated by the posterior distribution plot in Figure~\ref{fig:demo_posterior}. Furthermore, these functionalities are accessible via a simple YAML configuration file, with the corresponding API briefly outlined in \url{https://github.com/caoxiaoyue/TinyLensGpu/tree/main/paper/demo}.

\subsubsection{Statistical Results on Mock Lenses}
\label{sec:stat_mock}
We adopt a model configuration that employs a singular isothermal ellipsoid (SIE) profile for the lens galaxy's mass and a S\'ersic profile for the light distributions of both the lens and source galaxies. We apply this configuration to model 1,000 mock lenses. In principle, because our modelling setup exactly matches that used to generate the mock data and is free from systematic errors, all lens parameters should be accurately recovered provided that the non-linear search effectively explores the complex parameter space. Since the primary challenge in automatic lens modelling is ensuring a thorough exploration of the parameter space, the results of this test offer valuable insights into the robustness of our pipeline and the sampling reliability of the \texttt{nautilus-sampler}.

In Figure~\ref{fig:mock_summary}, we compare the lens modelling results from the 1,000 mock lenses with the ground truth. We assess both the lens mass parameters (e.g., Einstein radius $\theta_{E}$, axis ratio $q_l$, and position angle $\phi_l$) and the source light parameters (e.g., half-light radius $r_s$, axis ratio $q_s$, and apparent magnitude $m_s$). Our models generally agree well with the ground truth, with most data points clustering around the 1:1 relation indicated by the red solid line. To statistically quantify the accuracy of the lens modelling, we compute the Normalised Median Absolute Deviation (NMAD):
\[
\text{NMAD} = 1.48 \times \text{median} \left( \left|\zeta_i - \text{median}(\boldsymbol{\zeta})\right| \right),
\]
where $\boldsymbol{\zeta}= \{\zeta_1, \zeta_2, \dots, \zeta_n\}$ and $\zeta_i$ denotes the deviation for the $i^{th}$ sample, defined either as the relative deviation $\left(\frac{\mathrm{Model}-\mathrm{True}}{\mathrm{True}}\right)$ or as the absolute difference ($\mathrm{Model}-\mathrm{True}$). Our analysis reveals that the typical relative deviations for $\theta_{E}$ and $r_s$ are approximately $0.2\%$ and $0.5\%$, respectively. The typical absolute deviations for $q_l$ and $q_s$ are $0.010$ and $0.013$, respectively. Furthermore, the position angle $\phi_l$ is measured with an average deviation of $1.27^\circ$, and the apparent magnitude $m_s$ with a deviation of $0.031$ magnitudes. Overall, these results demonstrate that the lens model parameters are statistically well recovered, highlighting the baseline performance of \texttt{TinyLensGpu} for automated modelling of large lens datasets.

However, we also identify several “catastrophic outliers” where the model fails to recover the ground truth accurately. We define these outliers as samples with deviations exceeding four times the typical NMAD scatter. Approximately $5\%$ of the cases fall into this category, mainly due to sampling failures. We discuss these cases in detail in Section~\ref{sec:discuss_failure}.

\subsubsection{Performance Benchmarking}
\label{sec:perform_bench}
\begin{figure*}
	\includegraphics[width=\textwidth]{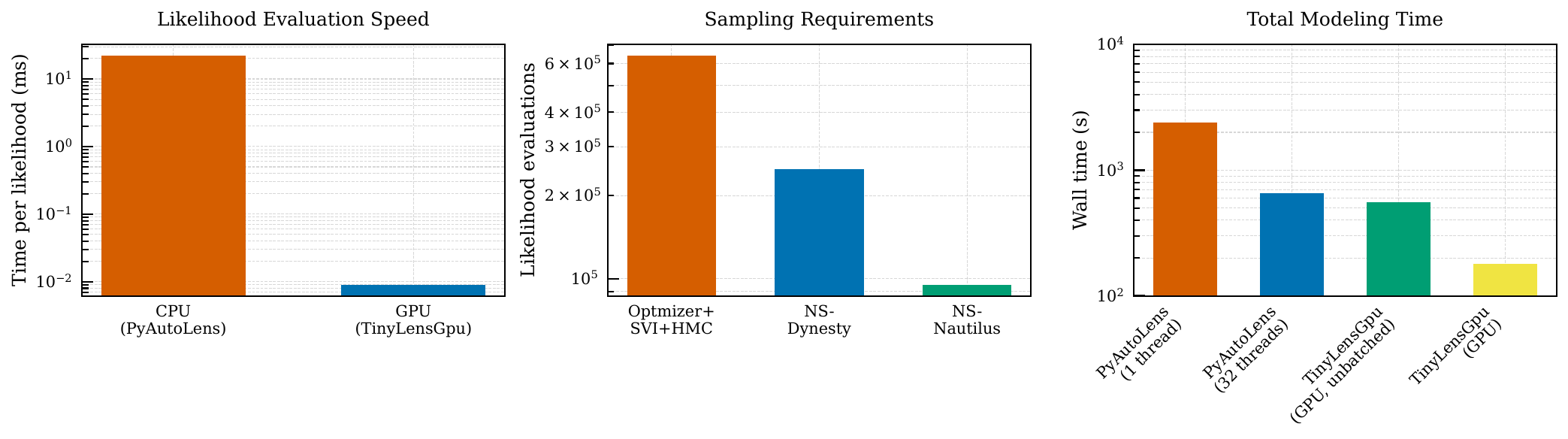}
        \caption{Performance comparison of \texttt{TinyLensGpu} with existing lens modelling tools, demonstrating improvements in likelihood evaluation speed, sampling efficiency, and overall modelling time. Left: \texttt{TinyLensGpu} achieves an approximately 2,000-fold speedup in likelihood evaluation time compared to the CPU-based \texttt{PyAutoLens}. Middle: The \texttt{nautilus-sampler} reduces the number of likelihood evaluations required for parameter space sampling from approximately one million (using an optimiser to find the maximum likelihood solution, followed by SVI and HMC for sampling) to approximately one hundred thousand. Compared to the \texttt{dynesty} nested sampler, \texttt{nautilus-sampler} typically achieves a three-fold improvement in sampling efficiency. Right: For a typical galaxy-galaxy strong lens system with an image size of $200 \times 200$ pixels, \texttt{TinyLensGpu} running on a GPU without batching achieves comparable speed to \texttt{PyAutoLens} utilising 32 CPU threads. Batching likelihood evaluations with \texttt{TinyLensGpu} can further enhance the modelling speed by a factor of three compared to the non-batched mode.}
    \label{fig:speed_comparison}
\end{figure*}

We benchmark the performance of \texttt{TinyLensGpu} against \texttt{PyAutoLens}, one of the most widely used and highly optimised CPU-based lens modelling codes. Our benchmark employs simulated images of a typical ring-like galaxy-galaxy strong lens, designed to replicate observations from the HST. The model components include a singular isothermal ellipsoid (SIE) with external shear for the lens mass and a Sérsic profile for both the lens and source light, where the surface brightness associated with the light profiles is solved linearly. As shown in Figure~\ref{fig:speed_comparison}, using GPU acceleration and batched likelihood evaluations, \texttt{TinyLensGpu} performs single likelihood evaluations (that is, one forward image simulation) approximately 2000 times faster than \texttt{PyAutoLens}. Additionally, \texttt{TinyLensGpu} employs the \texttt{nautilus-sampler}, which integrates a neural network to efficiently propose new sampling points. This reduces the number of likelihood evaluations required by a factor of 3 to 6 compared to established methods such as the \texttt{dynesty} nested sampler or hybrid approaches combining gradient descent optimisation, stochastic variational inference, and HMC \citep[e.g.][]{Gu2022}. Overall, \texttt{TinyLensGpu} achieves at least a tenfold speedup over conventional CPU-based codes, reducing the modelling time per lens from days to approximately 3 minutes, with even greater efficiency gains when processing larger datasets, such as lensing images of group-scale lenses. GPU-based codes such as \texttt{gigalens} require approximately 3 minutes for optimisation, 5 minutes for SVI, and 23 minutes for HMC for the benchmark lens system\footnote{While \cite{Gu2022} report a modelling time of approximately 100 seconds per lens using \texttt{gigalens}, our tests required around 30 minutes. Although differences in nonlinear search settings may contribute to this discrepancy, the primary factor is the hardware: \cite{Gu2022} used four H100 GPUs, whereas our tests employed a single RTX 4060 Ti, which has significantly lower computational power.}. Consequently, \texttt{TinyLensGpu} also exhibits superior modelling speeds, largely owing to the efficient sampling provided by the \texttt{nautilus-sampler}.

However, it is important to note that despite achieving likelihood evaluation speeds approximately 2000 times faster than conventional CPU-based code, this acceleration does not translate proportionally to overall modelling speed. This limitation stems from two factors: first, only the likelihood evaluation runs on the GPU while \texttt{nautilus-sampler} remains CPU-bound; second, frequent data transfers between CPU and GPU create additional overhead. We discuss potential strategies to mitigate this limitation by fully porting \texttt{nautilus-sampler} to the GPU in Section~\ref{sec:discuss_gpu_sampling}.

\subsection{Validation on Real Lensing Data}
\label{sec:res_real}
\begin{figure*}
	\includegraphics[width=\textwidth]{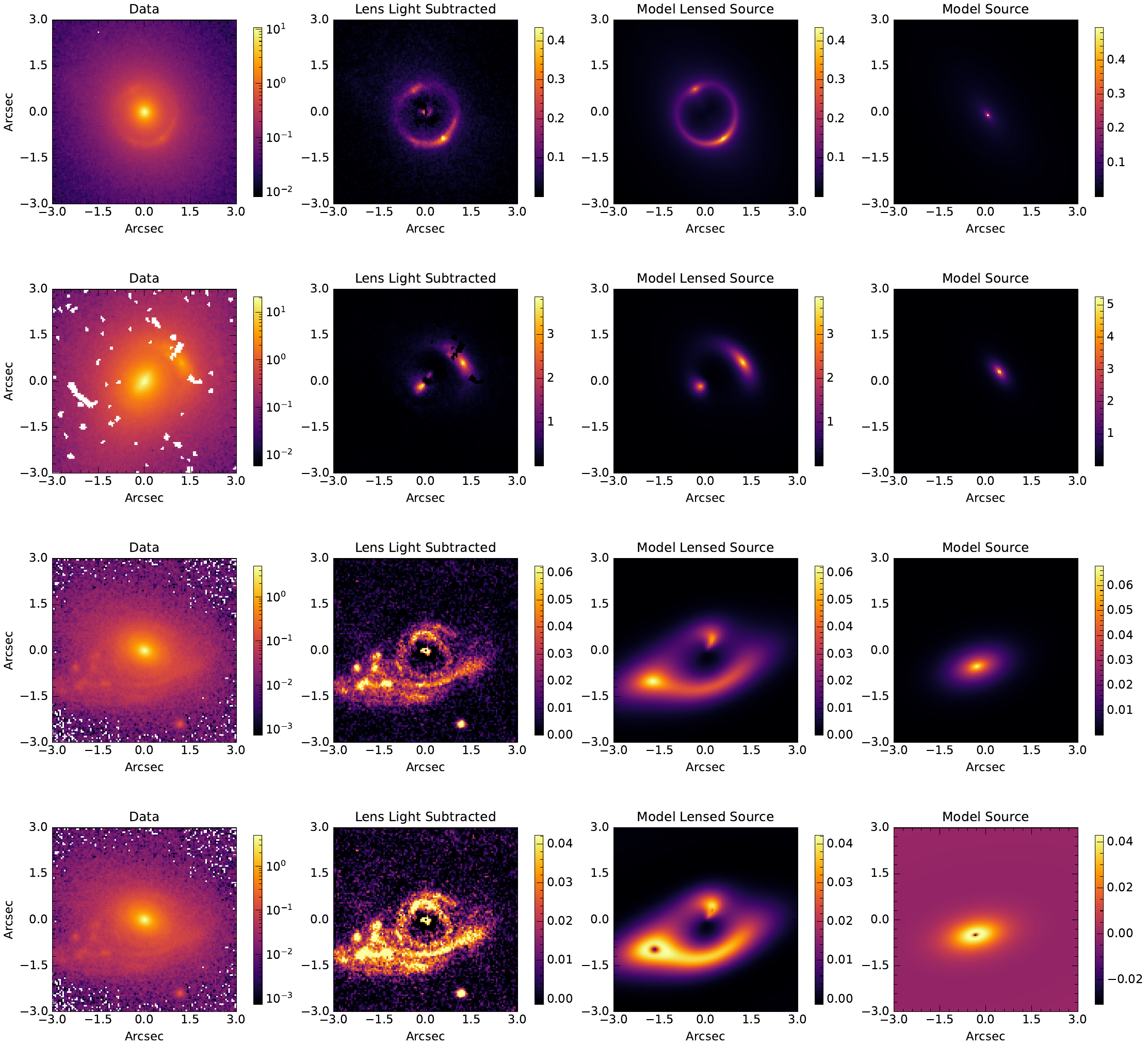}
        \caption{Lens modelling results for three representative systems from the SLACS dataset (from top to bottom: J0029-0055, J0044+0113, and J0157-0056). Each row presents, from left to right, the observed image, the residual image after subtracting the model-derived lens light, the reconstructed lensed source image, and the corresponding unlensed source image in the source plane. The bottom row illustrates the modelling result for the third lens, where non-negative linear least-squares fitting was not used to solve intensities. As a result, unphysical negative intensities appear in certain light profile components, creating holes in the light distribution.}
    \label{fig:slacs_example}
\end{figure*}

\begin{figure}
	\includegraphics[width=\columnwidth]{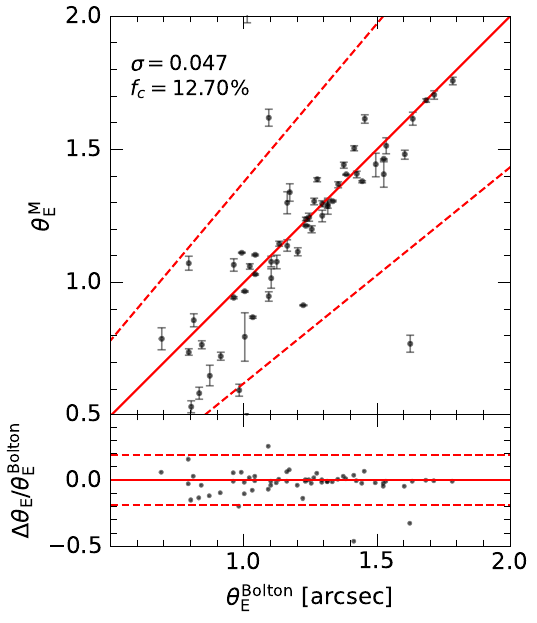}
        \caption{Similar to Figure~\ref{fig:mock_summary}, this figure compares the Einstein radius measurements from this work with those reported by \citet{Bolton08}.}
    \label{fig:slacs_summary}
\end{figure}
We analyse 63 Grade-A SLACS lenses using a uniform lens modelling pipeline. The image data were processed using the custom pipeline described in \cite{Bolton08}. The lens mass distribution was modelled as a singular isothermal ellipsoid (SIE) with external shear, while the light profiles of both the lens and source were represented by two Sérsic models. To mitigate degeneracies during sampling, the two Sérsic components for the lens and source shared a common centre and ellipticity, with their Sérsic indices fixed at 1 and 4, respectively. Figure~\ref{fig:slacs_example} presents the modelling results for three randomly selected lens systems. Our simplified parametric lens model accurately reproduces the global morphology of the observed lensing images. A quantitative comparison between our results and those from \cite{Bolton08} in measuring the Einstein radius is shown in Figure~\ref{fig:slacs_summary}. Our measurements are in excellent agreement with those of \cite{Bolton08}, exhibiting a typical scatter of $\sim 5\%$. This discrepancy aligns with the systematic errors reported in \cite{Bolton08}, particularly considering differences in model assumptions\footnote{Unlike \citet{Bolton08}, who employed a single SIE for the lens mass model, our analysis incorporates external shear.  Furthermore, our source model constrains the two Sérsic profiles to share a common centre and ellipticity, effectively representing a single light blob on the source plane, whereas \citet{Bolton08} allowed multiple free Sérsic components to model multiple light blobs.}. However, approximately $10\%$ of the systems exhibit significant discrepancies, where our uniform lens modelling fails to recover the correct results. These catastrophic outliers are discussed in detail in Section~\ref{sec:discuss_failure}.

%% file: discuss.tex
\section{Discussion}
\label{sec:discuss}
Section~\ref{sec:discuss_failure} summarises the failures encountered during the automated analysis of 1,000 mock lenses and 63 real HST lenses, offering valuable insights into uniform lensing analysis in the era of big data. Section~\ref{sec:discuss_gpu_sampling} examines the primary performance bottleneck in the current version of \texttt{TinyLensGpu} and proposes strategies for mitigation. Finally, Section~\ref{sec:discuss_advance_model} outlines future improvements to \texttt{TinyLensGpu}, with a focus on incorporating more advanced models for lens and source light.

\subsection{Failures in Automated Lens Modelling}
\label{sec:discuss_failure}
Approximately 5\% of the 1000 mock lenses and 10\% of the 63 real HST lenses failed to yield successful lens models. This failure resulted from the sampler's inability to identify a model solution that accurately captured the global morphology of the lensing images. A visual inspection of these catastrophic failures reveals two primary causes:
\begin{enumerate}
\item A low signal-to-noise ratio (SNR) in the lensing image results in residuals from an inaccurate lens model that remains indistinguishable from noise.
\item Some lensed systems contain a faint counter-image among the multiple-image pairs. Since this counter-image contributes minimally to the likelihood function, the sampler frequently overlooks it, leading to biases in the reconstruction of both the lens mass and the source light. This issue is particularly pronounced in real HST data, where the light from the lens galaxy cannot be perfectly represented by one or a few Sérsic models\footnote{Examples of SLACS systems for which the automated modelling fails are provided in Appendix~\ref{sec:appdx_A}.}. Residuals from imperfect lens light models may then be misinterpreted as lensed features, further complicating the modelling process.
\end{enumerate}

Since these catastrophic failures mainly stem from unsuccessful sampling, we investigate whether enforcing a more thorough exploration of the parameter space could yield unbiased modelling results. Specifically, we increase the number of live points in the \texttt{nautilus-sampler} from 300 to 500, 1000, and 3000, extending the modelling time from approximately 3 to 30 minutes. However, even with this more exhaustive sampling strategy, catastrophic failures persist. Similarly, experiments with the \texttt{dynesty} sampler and the gradient-based optimiser produce comparable results. In certain cases, the non-linear search struggles to fully explore the parameter space, making accurate modelling difficult.

To mitigate these catastrophic failures, imposing informative priors on the lens mass ellipticity or incorporating prior knowledge of the positions of the lensed images \citep{Nightingale2018, Cao2022, Etherington2022} proves beneficial. This suggests a general strategy for reducing catastrophic failures in uniform lens modelling: (1) leveraging neural network-based lens parameter predictors to provide informative priors on model parameters or (2) utilising multi-band colour information or neural networks to pre-detect faint counter-images \citep[e.g.][]{Shajib2025}, guiding the sampler away from local optima. Consequently, a hybrid approach that combines machine learning with forward simulation techniques could enhance automated lens modelling \citep[e.g.][]{Pearson2021}, particularly in mitigating catastrophic outliers.

\subsection{Towards Fully GPU-based Sampling}
\label{sec:discuss_gpu_sampling}
A primary performance bottleneck in the current implementation of \texttt{TinyLensGpu} is the reliance on CPU-based \texttt{nautilus-sampler} for the sampling process. The multilayer perceptron, which generates new sampling points (live points), represents a computationally intensive component of \texttt{nautilus-sampler}. Performance could be significantly enhanced by accelerating this process with JAX to leverage GPU capabilities. Moreover, the nested sampling algorithm itself does not fundamentally preclude GPU execution or a JAX-based implementation. In fact, the feasibility of porting \texttt{nautilus-sampler} to JAX has been demonstrated by community-developed JAX-based nested sampling codes such as \texttt{jaxns} \citep{JaxNs_1, JaxNs_2}. We currently do not employ \texttt{jaxns} for sampling because, in our practical lens modelling tests, it becomes trapped in local optima more frequently than \texttt{nautilus-sampler}. Furthermore, \texttt{jaxns} often requires millions of likelihood evaluations to achieve convergence. Consequently, despite its capacity to utilise GPUs, \texttt{jaxns} did not outperform \texttt{nautilus-sampler} in terms of overall modelling speed or sampling reliability. We hypothesise that this behaviour arises because the method \texttt{jaxns} employs to propose new sample points (a modified version of slice sampling) is less effective than the approach used by \texttt{nautilus-sampler} (which utilises a multi-layer perceptron) in navigating the noisy lens modelling parameter space. We plan to explore porting the \texttt{nautilus-sampler} to JAX to fully exploit the computational potential of the GPU.

We do not currently incorporate gradient information from JAX's automatic differentiation in our sampling process, as the neural network–based \texttt{nautilus-sampler} effectively addresses the challenges of parameter inference in automated lens modelling when initialised with a non-informative prior. Moreover, the conventional approach—optimizing to obtain an initial parameter estimate, applying stochastic variational inference, and then using HMC to derive the complete posterior—does not surpass the \texttt{nautilus-sampler} in terms of the number of likelihood evaluations required or overall modeling speed (see Section~\ref{sec:perform_bench}). Therefore, we continue to use the \texttt{nautilus-sampler} in this work. Nevertheless, gradient information remains crucial, especially when fitting complex models with hundreds or even thousands of free parameters.

\subsection{Advanced Models for Lens and Source Light}
\label{sec:discuss_advance_model}
We currently use analytical Sérsic models to represent the light distributions of both lens and source galaxies. While this simplified approach provides catalogue-level results and captures key lensing properties—such as the Einstein radius and the intrinsic source size and magnitude—it lacks the complexity required for detailed analyses. For instance, studies aiming to separate dark matter and luminous components in lens galaxies \citep{Nightingale2019} require more advanced models that better capture the intricate morphology of both lens and source light distributions. To address these limitations, we have implemented the Multiple Gaussian Expansion (MGE) model, which can describe arbitrary radial profiles and isophote twists in lens galaxies\footnote{Example scripts for modelling galaxy-galaxy strong lenses using the MGE model are available at \url{https://github.com/caoxiaoyue/TinyLensGpu/tree/main/paper/demo/lens_src_mge}.} \citep{He2024_MGE}. Additionally, we are developing pixelated source models \citep{Warren2003, Suyu2006, Nightingale2015} capable of representing highly irregular source morphologies \citep[e.g.][]{Shu16}, which will soon be integrated into \texttt{TinyLensGpu}.

%% file: summary.tex
\section{Summary}
\label{sec:summary}
We have developed \texttt{TinyLensGPU}, a lens modelling software designed to address the computational challenges of analysing large lensing datasets from ongoing and upcoming space telescope surveys, including Euclid, CSST, and Roman. \texttt{TinyLensGPU} is implemented in Python and utilises JAX to accelerate likelihood calculations by leveraging GPU computing. This optimisation reduces the computation time for simulating a single lensing image from 10–100 milliseconds to approximately 10 microseconds, achieving a speed improvement of nearly three orders of magnitude. Additionally, the \texttt{nautilus-sampler} has been integrated to enhance sampling efficiency. This sampler employs a multilayer perceptron to propose new sampling points, reducing sensitivity to local optima and enabling a thorough nonlinear search with significantly fewer likelihood evaluations (approximately $10^4$ steps compared to $10^5$–$10^6$ steps in traditional methods). By combining JAX-accelerated likelihood computations with the efficient \texttt{nautilus-sampler}, the pipeline achieves an order-of-magnitude speedup over conventional modelling tools for mock CSST strong lenses, reducing the modelling time to approximately three minutes. Consequently, \texttt{TinyLensGPU} is well-suited for applications requiring rapid lens modelling, such as providing timely information for follow-up time-delay observations of lensed supernovae.

Evaluations on 1,000 mock lenses designed to emulate CSST observation and 63 real SLACS lenses imaged by HST demonstrate that \texttt{TinyLensGPU} reliably recovers key lensing parameters, such as Einstein radius and intrinsic size and magnitude of the source, in most cases. However, approximately $5\%-10\%$ of cases become catastrophic outliers, failing the lens modelling and producing incorrect solutions. Further analysis indicates that these outliers may result from faint counter-images with low signal-to-noise ratios, which contribute minimally to the likelihood and are therefore often overlooked by the sampler. Moreover, residuals from imperfect lens light modelling can be misinterpreted as lensed source features, leading to failed models. Incorporating additional prior information from machine learning techniques, such as constraints on lens mass ellipticity or the positions of the lensed images (particularly faint ones), can mitigate these issues by guiding the sampler away from local optima. Thus, integrating machine learning with forward-simulation-based lens modelling is crucial for further reducing the occurrence of outliers in automatic lens modelling.

\texttt{TinyLensGPU} scales efficiently with increasing dataset sizes due to the GPU's ability to handle large-array computations. Next-generation telescopes, such as the Extremely Large Telescope \citep[ELT,][]{ELT07}, will offer resolutions up to an order of magnitude higher, generating imaging datasets for lens modelling that are hundreds of times larger. Unlike conventional CPU-based lens modelling codes, which may require weeks or months to process such extensive images, \texttt{TinyLensGPU} can complete the analysis in approximately 10 minutes. Although initially designed for galaxy-galaxy strong lenses, its capability to handle large datasets makes it well-suited for modelling group- and cluster-scale lenses. Such systems present significant challenges for traditional CPU-based approaches due to the enormous image arrays associated with extended lensed arcs. With the advent of high-resolution space telescopes, GPU-accelerated lens modelling software is expected to become the new community standard. Furthermore, the flexible architecture of \texttt{TinyLensGPU} enables future enhancements. We plan to incorporate more sophisticated models—such as the pixelated source model—to further expand its applicability and improve modelling accuracy.

%% file: appdx_A.tex
\section{Examples of Failure Modelling in the SLACS Sample}
\label{sec:appdx_A}
\begin{figure*}
	\includegraphics[width=\textwidth]{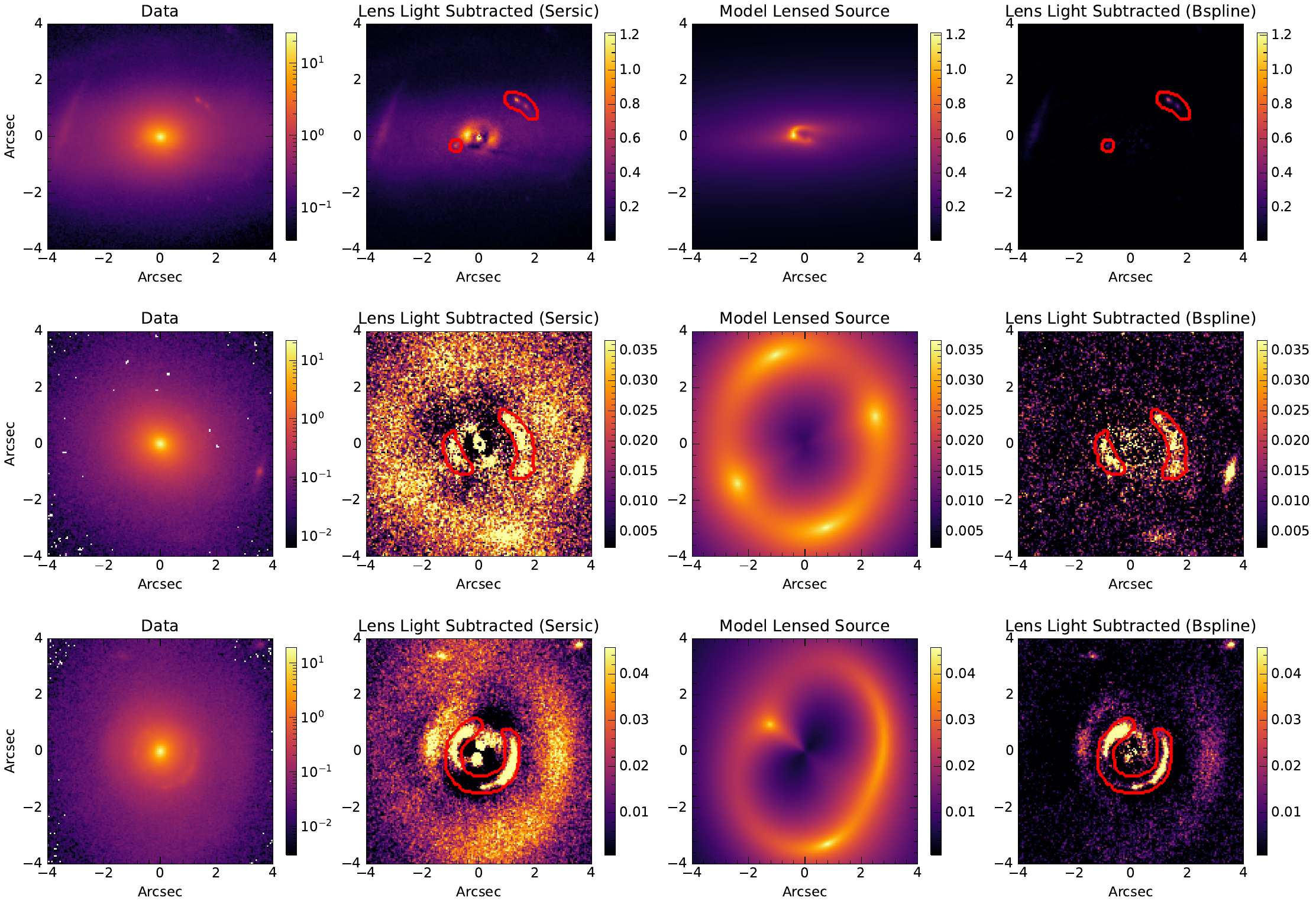}
        \caption{Examples of three SLACS systems (top to bottom: J0841+3824, J1016+3859, and J1153+4612) for which our automated lens modeling fails. In each row, panels show (from left to right): (1) the observed image; (2) the residual image after subtracting the lens light modeled with two concentric S\'ersic components; (3) the reconstructed lensed source; and (4) the lens light subtracted image from \citet{Bolton08}, who used a more flexible B-spline model to fit the lens light. The red curves mark the region of the observed lensed source that the lens model was supposed to fit.}
    \label{fig:slacs_failures}
\end{figure*}

For a more intuitive impression of the SLACS lens systems that our automated modelling fails, we present three such examples in Figure~\ref{fig:slacs_failures}: J0841+3824 (top row), J1016+3859 (middle row), and J1153+4612 (bottom row). In each row, from left to right, we display (i) the observed lensing image; (ii) the residual image obtained by subtracting the lens light with two concentric S\'ersic models and the reconstructed lensed image, both produced by our automated lens-modelling pipeline; and (iii) the lens-light-subtracted images of \citet{Bolton08}, who employ a more flexible B-spline model capable of fitting boxiness, diskiness and arbitrary radial profiles of the lens galaxy light, thereby modelling the lens-light distribution more accurately\footnote{Despite its flexibility, the B-spline fitting procedure involves manually choosing spline break points, masking out the lensed emission, adding optional angular complexity and an iterative modelling routine, all of which are difficult to automate \citep{Bolton06, Bolton08}.}. The red curves delineate the regions of the observed lensed image that the lens model was supposed to fit. In J0841+3824, the faint counter-image near the lens centre may be ignored by the sampler due to its negligible contribution to the likelihood. Owing to the complexity of the lens-light distribution in this system, two S\'ersic models cannot fully recover it, resulting in significant residuals around the lens centre after subtraction. Our automated pipeline appears to interpret these subtraction residuals as part of the lensed source and attempts to fit them with the source model, thereby neglecting the actual lensed emission. Similar phenomena are observed in J1016+3859 and J1153+4612, where the source model attempts to fit the outer ring-like residuals left by imperfect lens light subtraction. In future work, we aim to incorporate additional prior information from machine learning based techniques—such as the Einstein radius and ellipticity of the lens mass \citep[e.g.][]{Pearson2021} and the positions of lensed images \citep[e.g.][]{Shajib2025}—to guide the sampler away from local maxima and thus further reduce the occurrence of catastrophic modelling failures.

